# Sensitively Temperature-Dependent Spin Orbit Coupling in SrIrO$_3$ Thin Films

Lunyong Zhang[1], Y. B. Chen[*,2], Jian Zhou[1], Shan-Tao Zhang[1], Zheng-bin Gu[1], Shu-Hua Yao[1], and Yan-Feng Chen[**,1]

[1] National Laboratory of Solid State Microstructures & Department of Materials Science and Engineering, Nanjing University, Nanjing 210093 China

[2] National Laboratory of Solid State Microstructures & Department of Physics, Nanjing University, Nanjing 210093 China

Spin orbit coupling plays a non-perturbation effect in many recently developed novel fields including topological insulators and spin-orbit assistant Mott insulators. In this paper, strongly temperature-dependent spin orbit coupling, revealed by weak anti-localization, is observed at low temperature in 5d strongly correlated compound, SrIrO$_3$. As the temperature rising, increase rate of Rashba coefficient is nearly 30%-45%/K. The increase is nearly 100 times over that observed in semiconductor heterostructures. Microscopically, the large increase of Rashba coefficient is attributed to the significant evolution of effective Landé *g* factor on temperature, whose mechanism is discussed. Sensitively temperature-dependent spin orbit coupling in SrIrO$_3$ might be applied in spintronic devices.

## I. INTRODUCTION

Spintronics has been one central topic of condensed matter physics research in recent years. It proposes to use the "spin" of an electron to act the functions traditionally carried out by the "charge", so overcome the moore's law crisis. In the studies of spintronics, spin orbit coupling effect (SOC) attracts particular attentions because it is a fundamental mechanism to manipulate the spin degree of freedom[1]. A large number of researches have indicated that SOC plays crucial roles in the emergence of several non-trivial physical effects. For example, one is the topological insulator; the energy gap of several semiconductors such as HgTe is generated by SOC. It supports an edge state inside the band-gap[2]. Another case is the 5d Transition Metal Oxides Sr$_{n+1}$Ir$_n$O$_{3n+1}$, where SOC modifies electron state from $t_{2g}$ triplet to $J_{eff}$=1/2 and $J_{eff}$=3/2 multiplet. Fermi level crosses the $J_{eff}$ =1/2 doublet, electron filling in this state is half. Thus finite electron-electron interaction transforms the Sr$_2$IrO$_4$ (n=1) into a Mott insulator[3]. These discoveries suggest that SOC is not a perturbative effect in materials with heavy elements. Accordingly, we study the SOC effect in SrIrO$_3$ (n=∞), especially focusing on its temperature dependence. This is strictly substantial for spintronic device design and application and has been recently noticed showing unpredicted behaviors in a few of semiconductor heterostructure.

There are two considerations to choose SrIrO$_3$ as the study system. Firstly, being a three dimensional counterpart of Sr$_2$IrO$_4$ (quasi-two dimensional system), what is the effect of SOC on physical properties of SrIrO$_3$. Secondly, compared to the well studied SOC in semiconductor heterostructure, what is the special effect of SOC in oxides containing 5d transition metal elements? The researches of SOC so far were largely carried out in III-V and II-VI semiconductors. They have indicated that SOC can be introduced by structure inversion asymmetry (Rashba effect) as well as bulk inversion asymmetry (Dresselhaus effect) besides by the intrinsic mechanism [4], and can be modulated by strain and electrical effect. Compared with semiconductors, the transition metal oxides with electron-correlation always demonstrate diversely nontrivial physical behaviors such as high temperature superconductivity, charge and spin density wave, colossal magnetoresistance[5, 6]. SOC and electron-correlation of SrIrO$_3$ are about 0.4 and ~0.5 eV, respectively. It is interesting to suppose the SOC behavior in SrIrO$_3$ may be different from that in semiconductors. Several works have stated that electron-correlation enhanced SOC can modify electronic structure dramatically [7-9].

In the present work, we studied the SOC by fitting of anti-localization effect in magnetoconductance behaviour, extracted temperature-dependent SOC strength. A large temperature dependent SOC effect was discovered. Furthermore, we interpret it based on the effective Landé $g$ factor variation with temperature. The origin of temperature-dependent Landé $g$ factor and the comparison between SOC effect in semiconductor heterostructure and in SrIrO$_3$ are also discussed.

## II. EXPERIMENTAL

SrIrO$_3$ films with two thicknesses were synthesized on (001)-SrTiO$_3$ substrates by pulsed laser deposition (PLD) with a Brilliant Nd:YAG Laser of 355 nm laser and 160 mJ pulse energy at 750℃ substrate temperature in 25Pa oxygen atmosphere. The laser pulse frequencies for different thicknesses film growth were 2Hz and 10/3 Hz, respectively. The film growth duration was 10min. Detailed synthesis process can be browsed in Ref.10. The substrates were supplied by the Shanghai Daheng Optics and Fine Mechanics Co. Ltd. Their miscut angle is less than 0.5°. Before films synthesis, the STO substrates were treated for preparing a TiO$_2$ termination surface through the wide applied method discovered by Kawasaki *et.al* [11].

The x-ray diffraction (XRD) was carried out by a 2.4 kW Rigaku Rota flex X-ray diffractometer with Cu-kα ray. The film surface morphologies were recorded by an Asylum cypher atomic force microscopy (AFM). The film thicknesses $d$ were calibrated to be about 4nm and 7nm through a Tecnai F20 transmission electron microscopy (TEM). Electron transport measurements were carried out on a physical property measurement system (PPMS, Quantum Design) through the four-probe method

## III. RESULTS AND DISSCUSSIONS

The AFM image (Fig.1b) shows a typical smooth surface morphology with step growth mode for the 4nm SrIrO$_3$ films. The XRD pattern (Fig.1a) further indicates that both of the 4nm and 7nm films are orthorhombic phase SrIrO$_3$ (there was tiny monoclinic phase SrIrO$_3$ in the 7nm film since strain relaxation), which is consistent with our previous systematic research on the microstructure of SrIrO$_3$ film on (001)-SrTiO$_3$ substrate through TEM [10] and the reports of Kim [12].

The 7nm film shows obvious metal-insulator transition with transition temperature around 28K (Fig.2a), indicating quantum conductivity correction effects have emerged. Weak localization and electron-electron/or electron-boson interaction could be ascribed to the quantum conductivity correction mechanisms and they can be considered together through a temperature-dependent resistance model [13]

$$\rho(T) = \frac{1}{\sigma_0 + a_1 T^{p/2} + a_2 T^{1/2}} + bT^q \qquad (1)$$

where $\sigma_0$ labels the classical temperature-independent Drude conductivity, $a_1$ and $a_2$ respectively account for the weak localization (WL) and the electron-electron interaction (EEI) mechanism. $p$ describes weak localization effects, $p=2$ implies the dominance of electron-electron interaction and electron-boson scattering is the dominator if $p=3$[13]. The term $bT^q$ comes from normal inelastic scattering resistance, for example electron phonon scattering; the value of $q$ depends on the scattering mechanism. As shown in Fig. 2a, the temperature-dependent resistance can be soundly fitted in the temperature range above ~10K by Eq.1 with $p\approx3$, $q\approx0.44$, $\sigma_0\approx647.58\Omega^{-1}cm^{-1}$, $a_1\approx26.03$ $\Omega^{-1}cm^{-1}K^{-3/2}$, $a_2\approx98.89\Omega^{-1}cm^{-1}K^{-1/2}$ and

$b \approx 1.16 \times 10^{-5} \Omega$ cm $K^{-0.44}$. Therefore, weak localization is responsible for the metal-insulator transition around 28 K, which is further confirmed by the magnetoconductance behavior described as follows. And since $a_2$ is ~4 times as large as $a_1$, the conduction should be dominated by EEI at low temperature.

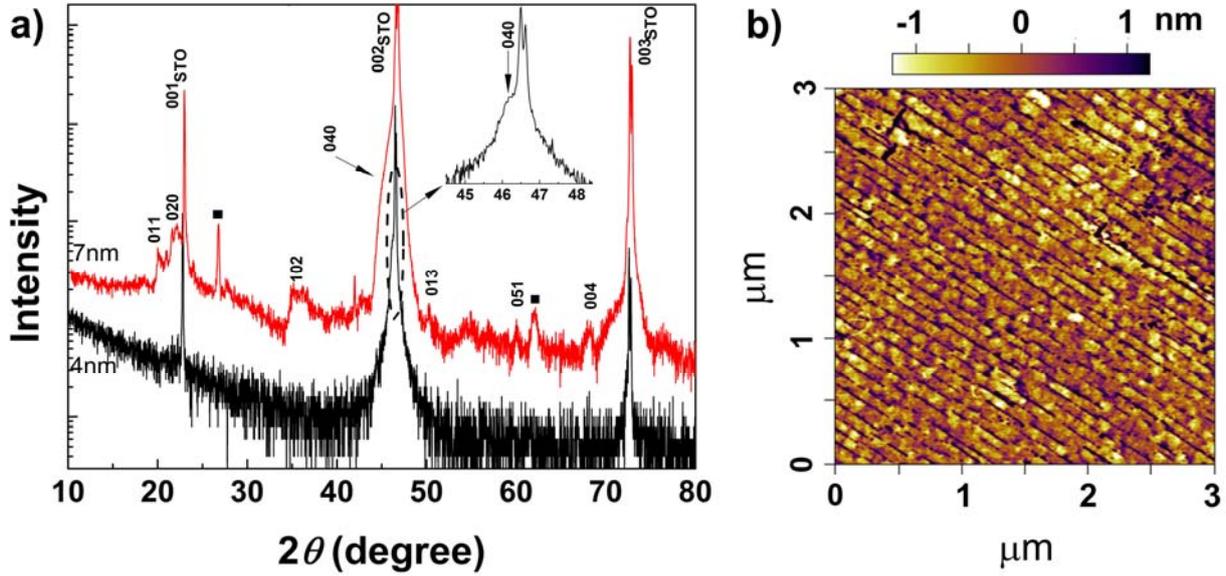

Fig.1 XRD patterns of the samples, a), and AFM image of the 4nm film as an example, b). The squares in XRD pattern mark the faint impure peaks belonging to a tiny of monoclinic $SrIrO_3$ phases.

To clarify the conduction mechanism below 10K, the typical variable range hopping (VRH) process in WL regime is considered based on the method proposed by Hill, et al. [14, 15]. Setting a quantity $w = -d\ln R / d\ln T$ associated with the general VRH expression $R = R_0 \exp(T_0/T)^{1/(D+1)}$, where $R_0$ and $T_0$ are two constants, and the $D$ is a constant indexing the hopping dimension. $D$ can be 2 and 3, denoting a two dimensional hopping and a three dimensional hopping, respectively. If conduction is dominated by VRH, $\ln w$ should be linearly dependent on $\ln T$. By linear fitting the $\ln w$ versus the $\ln T$ data, $D$ can be extracted from the slop of the linear fit, $1/(D+1)$. As briefly shown in Fig.2b, variable range hopping related conduction behavior was obvious. The hopping dimensional indexed by $D$ accurately equals two, implying that the film can be considered physically as a quasi two dimensional system [16] and is a great suitable platform for SOC study.

Differently, the 4nm film was insulative (Fig. 2a) and it entered the two dimensional VRH conduction regime at a temperature quite higher than that of the 7nm film, about 200K, suggesting the 4nm film was more two dimensionalized. Surprisingly, the dimensional parameter $D$ was near infinite according to the method mentioned above in the temperature

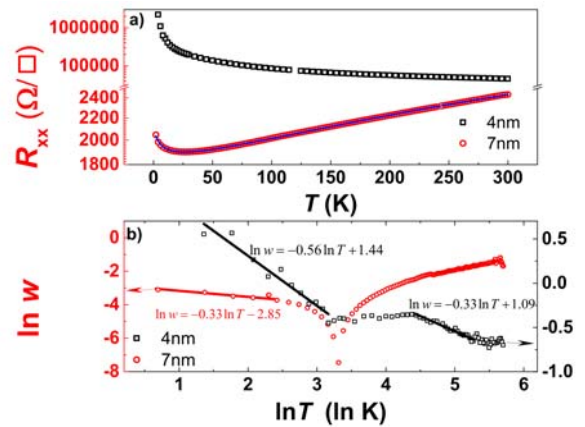

Fig.2 a) Temperature-dependent resistance curves of the 4nm and the 7nm $SrIrO_3$ films and related fitting results. b) The transformed curves of temperature-dependent resistances according to the variable range hopping model (VRH) and the related Hill- Zabrodskii-Zinov'eva VRH data treatment method.

range about 80K-25K, and furthermore this behavior can not be removed by a magnetic field perpendicular to the film (B=3Tesla, the data was not showed). Then D became about unit, indicating the Efros-Shklovskii variable range hopping conduction (ES-VRH) associated with Coulomb gap[17]. Therefore, the 4nm film was of stronger electron-electron interaction than the 7nm film. This point also has been suggested by their electron density comparison which was lower about 3-4 orders for the 4nm film, measured through the standard Hall effect.

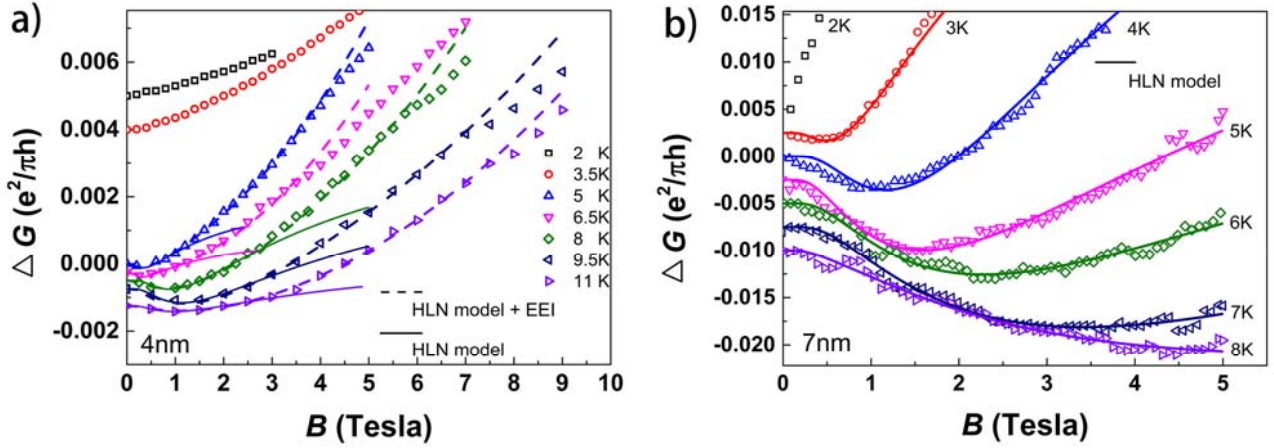

Fig.3 Magnetic-field dependent magnetoconductance traces of the 4nm a) and the 7nm b) $SrIrO_3$ films under different temperatures. Dots are the experimental data, the solid (through HLN model, Eq.2) and the dash curves (through HLN model with EEI correction) were the fitting results.

According to the well founded quantum correction theory of magnetoconductance, a system simultaneously exhibiting SOC and WL should have a concave-shaped conductance curve under a varied external magnetic field[18]. Fig.3 illustrates the magneto-conductance traces at several temperatures of the 4nm and the 7nm $SrIrO_3$ film samples. All of them showed characteristic concave shape at temperatures above 3.5K for the 4nm film and 2K for the 7nm film. The decrease of magneto-conductance in low field is caused by the SOC inducing weak anti-localization effect. After the conductance minimum, the weak anti-localization effect is overwhelmed by WL breakdown under large magnetic field. It results in the conductance increase. The field corresponding to the conductance minimum, $B_{min}$, is believed to approximately proportionate to the SOC strength[19], therefore a rough conclusion can be drawn that the SOC strength of the $SrIrO_3$ film in the investigated temperature range exhibits a positive correlation to temperature.

In theory, magnetoconductance with SOC and WL quantum correction in low field around $B_{min}$ could be formulated by the Hikami-Larkin-Nagaoka equation (HLN model)[18]

$$\frac{\Delta G_{SOC+WL}}{G_u} \propto \left\{ \psi\left(\frac{1}{2}+\frac{B_e}{B}\right) - \psi\left(\frac{1}{2}+\frac{B_i+B_{soc}}{B}\right) + \frac{1}{2}\left[\psi\left(\frac{1}{2}+\frac{B_i}{B}\right) - \psi\left(\frac{1}{2}+\frac{B_i+2B_{soc}}{B}\right)\right] \right\} \quad (2)$$

where $\psi(x)$ is the digamma function, $\Delta G = G(B)-G(B=0)$ and $G_u = e^2/(\pi h) \approx 1.2\times 10^{-5}$ S is a universal value of conductance. The $B_e$, $B_i$ and $B_{soc}$ are the equivalent fields of elastic scattering, inelastic scattering and the scattering induced by SOC, respectively. Their magnitudes measure the corresponding scattering strengths. All of them can be expressed as functions of their scattering lengths $l_o$ (o=e, i and soc), i.e. $B_o = \hbar/4el_o^2$ [20]. Besides, if the magneto-transport of the system is affected by EEI, when the condition $k_B T \tau_{tr}/\hbar \ll 1$ is satisfied (at 0.01 scale in our samples), the EEI correction to magnetoconductance can be expressed by a parabolic relation with the magnetic field[21] $\Delta G_{EEI}/G_u \propto B^2$, where $k_B$ and $\hbar$ are the Boltzman constant and the Plank constant respectively, $\tau_{tr}$ is transport relaxation time which can be calculated through the Hall mobility $\mu$ with relation $\tau_{tr}=\mu m_*/e$. Here $m_*$ refers the effective electron mass, it equals to $7m_0$ for $SrIrO_3$[22] ($m_0$ is the free electron mass). And for a system affected by mixed corrections with SOC+WL+EEI, $\Delta G=\Delta G_{SOC+WL}+\Delta G_{EEI}$.

We fitted the magnetoconductrance data shown in Fig.3 by both the HLN model and the HLN model adding EEI correction. It was founded from Fig.3a that EEI interaction was necessary for good fitting the data of 4nm film. The $B_{soc}$ fitted by two methods showed increasing difference with the temperature rising (Fig.5a). On the contrary as shown in Fig.3b, both of the two methods can soundly fit the data of 7nm film and there were no big differences in the $B_i$ and $B_{soc}$, especially in the $B_{soc}$ (Fig.4 and Fig.5a). Fig.4 gives out the fitted $B_i$ at different temperatures. The $B_i$ here was around 0.02-0.03T, leading an inelastic scattering length $l_i$ at a scale of 200nm. It is close to those of ultrathin bismuth film [23] and graphene [24] at mesoscopic scale. Given that a mesoscopic system is defined by the sample size dimension over the inelastic scattering length, our $SrIrO_3$ films were mesoscopic systems in the growth direction since the thickness were remarkably smaller than the inelastic scattering length. In a mesoscopic system, all the particle collisions would be mainly dominated by elastic scattering, the coherent backscattering and the interference of particles are enhanced [25]. This naturally explains the weak localization conduction features in the samples (Fig.2b).

The $B_i$ did not show obviously temperature dependence in the measuring temperature range (Fig.4), suggesting a temperature insensitive inelastic scattering length. This is rare in normal system like metals where inelastic scattering length conventionally bears $T^{-v}$ relation with temperature. According to the wave scattering theory, a temperature dependent inelastic scattering length is held for quantum wave which means diffusion transport for a electron system, and the inelastic scattering length is generally insensitive to temperature for a classic wave case such as the electron transport in weak localization state [25]. It is reasonable that the sample here demonstrates a temperature insensitive inelastic scattering length. Actually, a non-temperature dependent inelastic scattering length also has been observed in an epitaxial graphene showing weak localization behavior[24].

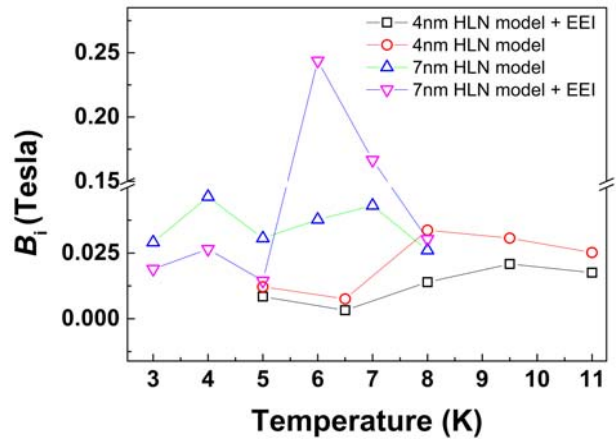

Fig.4 Equivalent fields of the inelastic scattering in the films obtained through the fitting of magnetoconductance traces shown in Fig.3

As for the SOC equivalent filed $B_{soc}$ (Fig.5a), it was of a order 0.01T in the 4nm film and 0.1T in the 7nm film, larger than those of many typical semiconductors (<10mT)[19, 26]. Of importance is that $B_{soc}$ distinctly increased with temperature. It leads Rashba SOC coefficient $\alpha$ of relative increasing rate ~30%/K in the 4nm film and ~45.4%/K in the 7nm film with temperature (Fig.5b) based on the expression $\alpha=m_*^{-1}\left(e\hbar^3 B_{soc}\right)^{1/2}$ [20] (our previous work showed a Rashba type SOC in $SrIrO_3$ film [27]). Temperature-dependent Rashba coefficient has also been observed in several semiconductor heterostructure/quantum well[29,30]. But it should be emphasized that the increasing rate shown above is nearly 100 times larger than that in semiconductor heterostructure (0.3-0.4%/K).[29, 30]

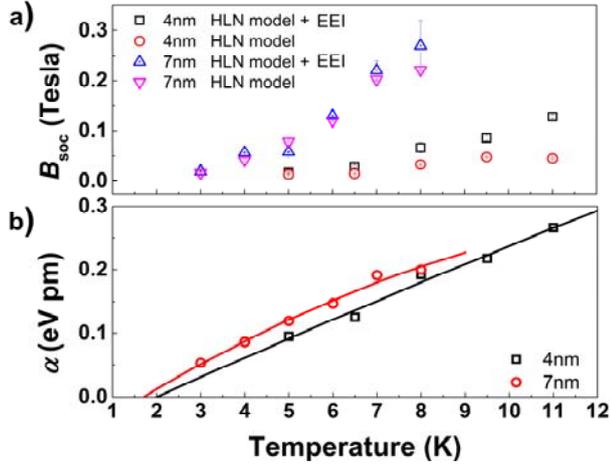

Fig.5. The spin orbit coupling equivalent fields, a), and the derived Rashba SOC coefficients, b), of the SrIrO$_3$ films, where the solid curves were obtained by fitting through Eq.4 with $g=1.0112-0.0056T$, $\varepsilon \approx 2.31\times 10^{15}$ V/m (4nm film), and $g=1.0568-0.033T$, $\varepsilon \approx 5.15\times 10^{14}$ V/m (7nm film), respectively.

To reveal the physical origin of temperature-dependent Rashba coefficient, we would like to analysis this phenomenon through the temperature dependence of Landé $g$ factor. By neglecting the band structure effect on the SOC, the Rashba coefficient could be expressed as

$$\alpha = g(1-g)\frac{\pi e \hbar^2 \varepsilon}{4 m_*^2 c^2} \quad (3)$$

Where $\varepsilon$ is asymmetric structure induced electric field, $c$ is the light speed [1]. Many experiments have proved in fact that the Landé $g$ factor is influenced by temperature, and in most cases shows near linear temperature dependence[28, 29]. Therefore, we have a modified Rashba coefficient expression by substituting $g=g_0+\lambda T$ into Eq.3 that is

$$\alpha = \left[g_0(1-g_0)+(1-2g_0)\lambda T - \lambda^2 T^2\right]\frac{\pi e \hbar^2 \varepsilon}{4 m_*^2 c^2} \quad (4)$$

As indicated in Fig.4c, the Rashba coefficient can be soundly fitted through Eq.4 with $g_0=1.0112$, $\lambda=-0.0056$ K$^{-1}$ and $\varepsilon \approx 2.31\times 10^{15}$ V/m for the 4nm film and with $g_0=1.0568$, $\lambda=-0.033$ K$^{-1}$ and $\varepsilon \approx 5.15\times 10^{14}$ V/m (greatly smaller than that of typical semiconductor systems, $\sim 10^{25}$ V/m [30]) for the 7nm film, suggesting that the proposal to interpret the temperature sensitive SOC by Landé $g$ factor variation with temperature would be feasible. Next, we further argue this interpretation through discussing the fitting results of Landé $g$ factor.

Regarding to the temperature dependence of Landé $g$ factor, many works have been done in several typical semiconductors like the GaAs and InSb during the past decades. It has been reported that rigid band model based on $k.p$ theory can not explain the temperature-dependent Landé $g$ factor. An accepted remedy on the discrepancy is based on band gap shrinking with the temperature increased. For example, Litvinenko et.al has explained the observed $g$ factor temperature dependence in both InSb and GaAs in the frame of $k.p$ theory by only taking the dilational variation of energy gap into account[31]. We therefore believe that the band gap shrinkage mechanism is basically applicable to our SrIrO$_3$ system. But we note that the fitting slop of the $g$ factor variation on temperature is nearly 1-2 order larger than that of several typical semiconductor systems (about 0.0004-0.0008 K$^{-1}$ [31, 32]). It infers that another factor impacting on band gap may exist besides the lattice expansion. This might be attributed to the electron-correlation in SrIrO$_3$ that is normally negligible in conventional semiconductors. It is well known that the Coulomb interaction could increase band gap and even split a band into two. But Coulomb interaction is weakened at high temperature because of enhanced screening effect.[36] Accordingly, we propose here a large shrinkage of band gap would be induced by the Coulomb interaction energy decrease when temperature is enhanced. Both lattice expansion and Coulomb interaction energy decrease induced band gap shrinkage lead to the large slop of the temperature-dependent $g$ factor in SrIrO$_3$.

Concerning $g_0$ representing the Landé $g$ factor at zero temperature. It has been founded that the regular $t_{2g}$ band with orbital angular momentum $L=1$ and spin angular momentum $S=1/2$ for $5d^5$-Ir$^{4+}$ would be split into effective total angular momentum $J_{eff}=1/2$ doublet band with high energy and $J_{eff}=3/2$ quartet band with low energy in the strong SOC limiting[33], the Landé g factor in this case is about 0.7 according to $g=1.5+[S(S+1)-L(L+1)]/2J(J+1)$ with $S=1/2$, $L=1$ and $J=1/2$. Contrarily, in the weak SOC limiting (without SOC splitting), it is 1.3 for the corresponding regular $t_{2g}$ band ($S=1/2$, $L=1$ and $J=L+S=3/2$). The $g_0=1.0112$ and =1.0568 for well fitting the Rashba SOC coefficient data of $SrIrO_3$ are between them, implying the $t_{2g}$ band just becomes a mixed state of the $J_{eff}=1/2$ band and $J_{eff}=3/2$ band instead of being fully split into two separated bands by SOC. As a result, the bandwidth of the mixed band in $SrIrO_3$ would be larger than that corresponding to the fully split bands as happened in SOC assistant Mott insulator $Sr_2IrO_4$, which is in consistent with the band structure of $SrIrO_3$ simulated from first principle method (LDA+U+SOC)[3] and naturally interprets the metallicity in $SrIrO_3$. This might suggests that the relative strong SOC is an indispensable ingredient to form SOC assistant Mott insulator in 5d iridium oxides. It also manifests that the fitting of α with Eq.4 is reasonable. Moreover, since with stronger Coulomb interaction, the mixed extent of the $J_{eff}=1/2$ band and $J_{eff}=3/2$ band of the 4nm film would be lower than that in the 7nm film, so the $g_0$ of the 4nm film should be closer to the g factor corresponding to the fully split case, 0.7, compared with the 7nm film as seen from the above Rashba coefficient fitting results through Eq.4.

Right now we have seen that temperature sensitive effective Landé g factor could attribute to the large variation of SOC in $SrIrO_3$ film when temperature is changed and the coulomb interaction weaken with temperature rising plays an essential role in causing the temperature sensitivity of the Landé g factor. In fact, it is well recognized that the effective mass also is affected by temperature, so similarly would has attribution to the Rashba coefficient variation on temperature according to Eq.3. However, this effect could be neglected comparing to the g factor attribution since the temperature dependence of the effective mass is generally much weaker than that of the g factor [34].

## IV. CONCLUSIONS

In summary we observed metal-insulator transition and spin orbit coupling induced weak anti-localization in $SrIrO_3$ films. Detailed analysis of temperature-dependent resistance and magnetoconductance suggested that Coulomb interaction was the dominating scattering mechanism at low temperature corresponding to the insulative regime with variable range hopping electron transport. And we found that the spin orbit coupling of $SrIrO_3$ films largely increases with temperature rising, leading to a near linearly temperature-dependent Rashba spin orbit coupling coefficient. Microscopically, it infers that the Landé g factor to be linear decreasing with temperature enhancement. Most remarkably, the relative increase rate of temperature-dependent Rashba coefficient was nearly two orders larger in magnitude than that of conventional III-V and II-VI semiconductor heterostructures, which could be applicable in temperature sensitive spintronic devices.

We'd like to acknowledge the financial support from the Nation Science Foundation of China (51032003, 50632030, 10974083, 51002074), the New Century excellent talents in University (NCET-09-0451), the China Postdoctoral Science Foundation (2013M530250) and the Jiangsu Province Postdoctoral Science Foundation (1202001C).

* ybchen@nju.edu.cn; ** yfchen@nju.edu.cn